\documentclass[twocolumn,showpacs,preprintnumbers,amsmath,amssymb,prl]{revtex4}

\usepackage{graphicx}
\usepackage{dcolumn}
\usepackage{bm}
\usepackage{tabularx}
\usepackage{amsmath}
\usepackage{textcomp}
\usepackage{upgreek}

\begin{document}

\title{High-pressure study of the Weyl semimetal NbAs}

\author{J. Zhang,$^{1}$ F. L. Liu,$^2$ J. K. Dong,$^{1,*}$ Y. Xu,$^{1}$ N. N. Li,$^2$ W. G. Yang,$^{2,\dag}$ and S. Y. Li$^{1,3,\ddag}$}

\affiliation{$^1$State Key Laboratory of Surface Physics, Department of Physics, and Laboratory of Advanced Materials, Fudan University, Shanghai 200433, P. R. China\\
$^2$Center for High Pressure Science and Technology Advanced Research, Shanghai 201203, P. R. China\\
$^3$Collaborative Innovation Center of Advanced Microstructures, Fudan University, Shanghai 200433, P. R. China}

\date{\today}

\begin{abstract}
We performed a series of high-pressure synchrotron X-ray diffraction (XRD) and resistance measurements on the Weyl semimetal NbAs. The crystal structure remains stable up to 26 GPa according to the powder XRD data. The resistance of NbAs single crystal increases monotonically with pressure at low temperature. Up to 20 GPa, no superconducting transition is observed down to 0.3 K. These results show that the Weyl semimetal phase is robust in NbAs, and applying pressure is not a good way to get a topological superconductor from a Weyl semimetal.

\end{abstract}

\maketitle


Following the intensive studies of topological insulators (TIs) \cite{XLQi1}, recently there are a lot of interests on three-dimensional (3D) topological semimetals \cite{XLQi2,Wanxg,Wenghm2,Young,Wang,Dai,Vafek,Turner,Wenghm,Hasan}, in which the conduction and valence bands touch at isolated points in the Brillouin zone, and the electrons have relativistic dispersion. Weyl semimetal (WSM) is one type of such materials, and it is characterized by the Weyl nodes, which always exist in pairs with opposite chirality \cite{XLQi2,Turner}. To get WSM, either time reversal or inversion symmetry needs to be broken \cite{Wang,Dai}. Though it is theoretically predicted to occur in strongly spin-orbit coupled systems, the existence of 3D WSMs are very rare in nature \cite{Wanxg,Wenghm2}. The topologically nontrivial 3D Dirac semimetal (DSM) is proposed to be protected by proper crystal symmetry when band inversion and time-reversal symmetry are present \cite{Wang,Dai}. Such protected Dirac nodes are four-fold degenerate states including spin degeneracy. In 3D DSMs, the degenerate Dirac point is composed of two overlapping Weyl points with opposite chirality \cite{Vafek}.

The 3D DSM phase has been experimentally discovered in Na$_3$Bi and Cd$_3$As$_2$ \cite{Liu1,Xu,Liu2,Neupane,Borisenko,Jeon,LPHe}, after the theoretical predictions \cite{Wang,Dai}. More recently, a family of noncentrosymmetric 3D WSM, the stoichiometric TaAs, TaP, NbAs and NbP with no inversion center in the crystal structure, was predicted \cite{Wenghm,Hasan}. A series of experiments have been performed on this family to search for the proposed WSM state \cite{Jias,Hasan2,Chengf,Dingh,Dingh2,Yanbh,FRonning,Hasan3}. As hallmarks of WSM state, both the Fermi arcs on the (001) surface and Weyl nodes in the bulk have been observed in TaAs and NbAs from ARPES measurements \cite{Hasan2,Dingh,Dingh2,Hasan3}. Extremely large magnetoresistance and ultra-high mobility was reported from transport measurements in TaAs, NbAs and NbP \cite{Jias,Chengf,Yanbh,FRonning}. Furthermore, negative magnetoresistance due to the chiral anomaly of Weyl fermions was also observed in TaAs \cite{Chengf}.

Starting from the topological semimetals, one may expect to obtain topological superconductors (TSCs) by carrier doping or applying pressure \cite{Dai,Dai2}. The TSCs have a full pairing gap in the bulk and gapless surface states consisting of Majorana fermions \cite{XLQi1}. Previously, a few candidates of TSC have been found by doping or pressurizing TIs \cite{YSHor,PDas,SSasaki1,SSasaki2,JLZhang,CZhang,Paglione,Matsubayashi,JZhu,PPKong}. For 3D DSM Cd$_3$As$_2$, superconductivity has been reported under high pressure, after a structural phase transition \cite{He}. The realization of pressure-induced superconductivity in these topological materials motivates us to study the high-pressure effects on WSMs. To our knowledge, there is still no pressure study on above-mentioned WSMs (TaAs, NbAs and NbP) so far. It will be very interesting to investigate whether pressure can induce superconductivity in them.

In this paper, we report the high-pressure powder X-ray diffraction and single crystal resistance measurements on the WSM NbAs. It is found that the crystal structure remains unchanged, and only the lattice parameters decrease slightly with increasing pressure. From the resistance measurements, no superconducting transition is observed down to 0.3 K and up to 20 GPa. These results show the robustness of the WSM phase in NbAs, and suggest that pressurizing a WSM may be not a good way to get a topological superconductor.

\begin{figure}
\includegraphics[clip,width=8.5cm]{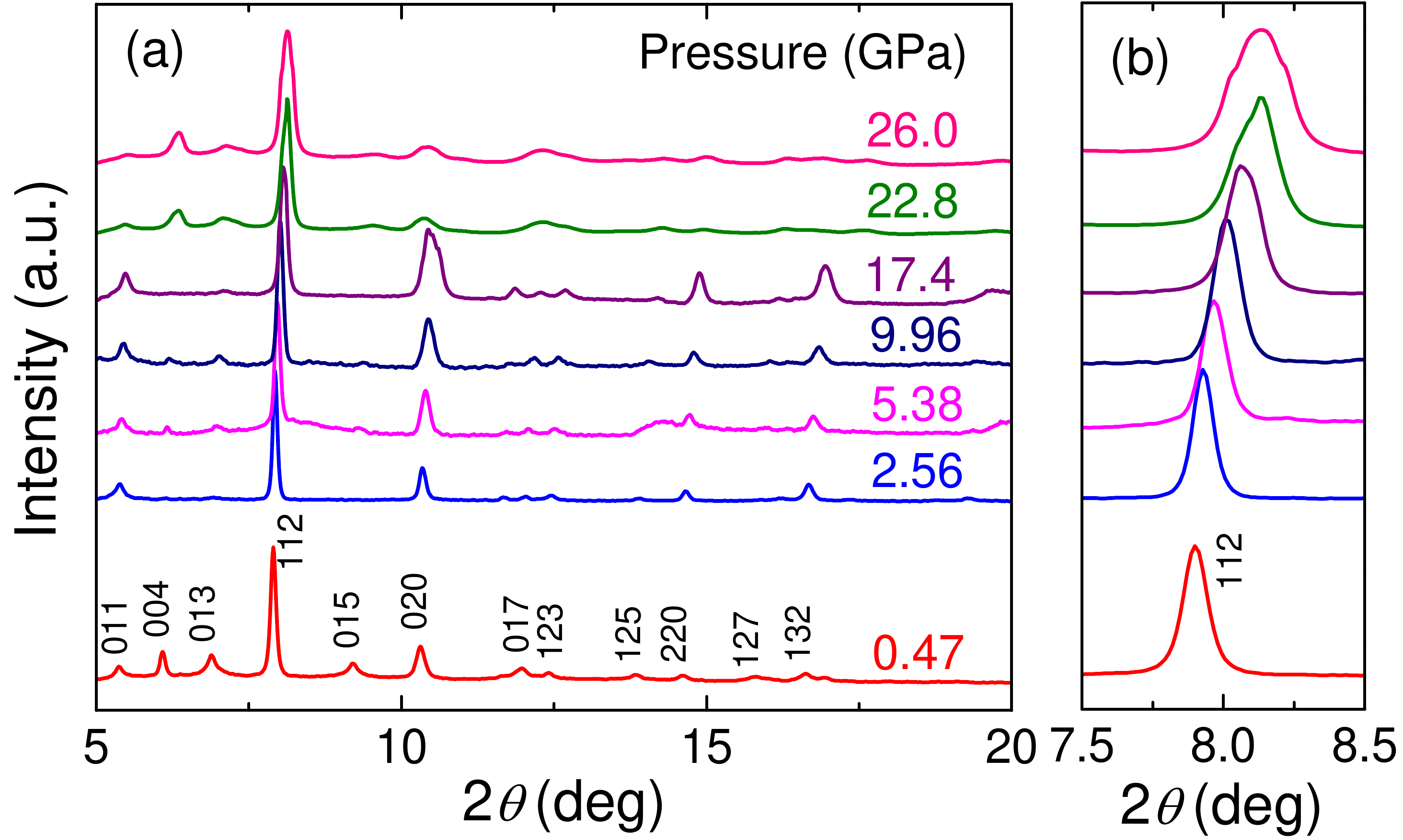}
\caption{(a) {\it In-situ} powder synchrotron XRD patterns of NbAs under various pressures up to 26 GPa at room temperature, with the X-ray wavelength $\lambda$ = 0.3100 {\AA}. (b) Pressure evolution of the (112) peak. The shift towards higher angle indicates the shrink of lattice under pressure.}
\end{figure}

\begin{figure}
\includegraphics[clip,width=8.5cm]{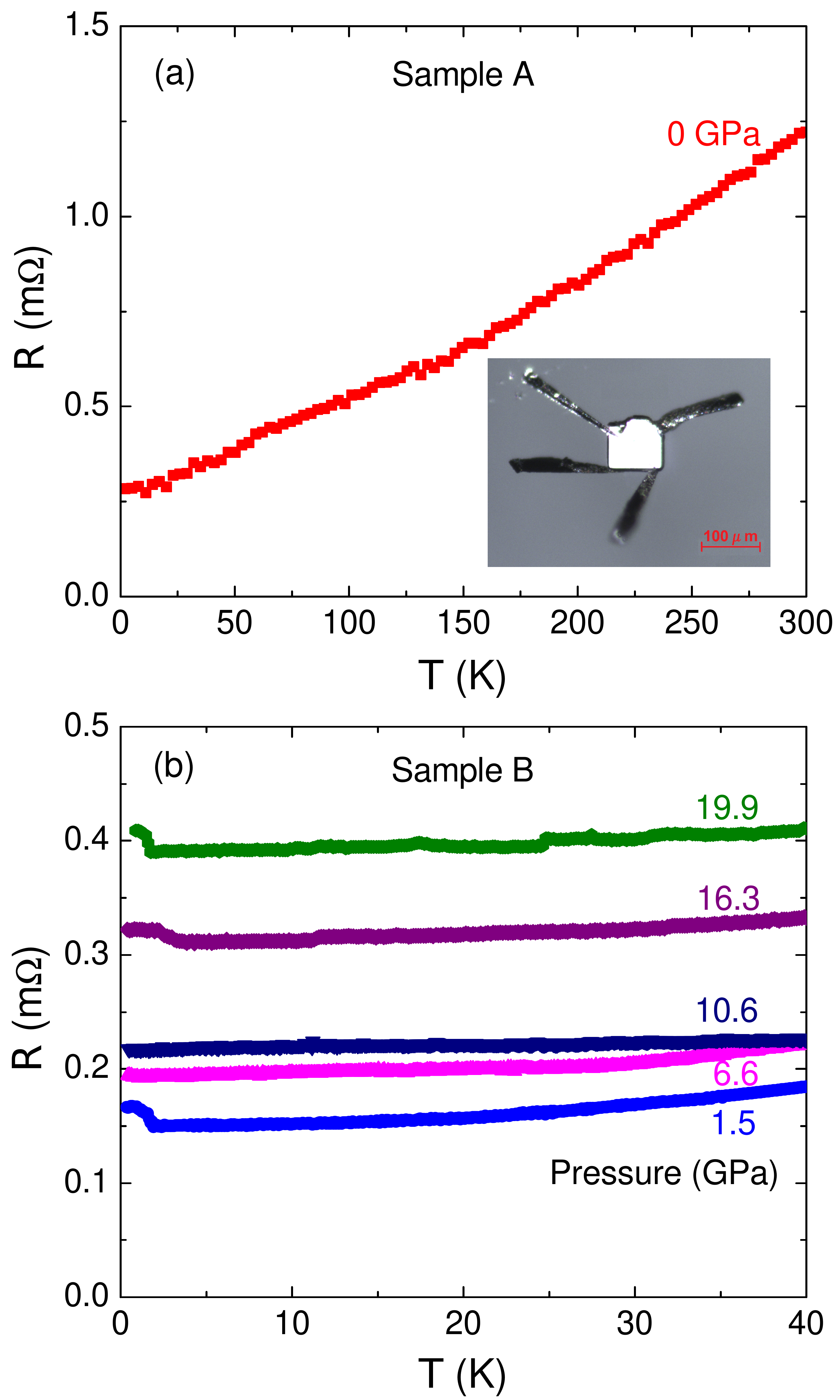}
\caption{The resistance of NbAs single crystal under various pressures. (a) $R(T)$ of sample A measured at ambient pressure. (b) $R(T)$ of sample B measured under pressures up to 20 GPa. Inset: the typical configuration of resistance measurement for NbAs single crystal. Four Pt electrodes were attached to the sample corners with silver epoxy.}
\end{figure}

The NbAs single crystals were grown using Sn flux method \cite{Wu}. After getting rid of the Sn flux, shinning single crystals with typical dimensions of 100 $\times$ 100 $\times$ 80 $\mu$m$^3$ were obtained. They are very stable in air and water. By using a Mao-Bell type diamond anvil cell (DAC) \cite{Yan}, the {\it in situ} powder synchrotron XRD experiments were carried out at High-Pressure Collaborative Access Team (HPCAT), at the Advanced Photon Source of Argonne National Laboratory. The powder was prepared by grinding several pieces of single crystals. For resistance measurements, the thickness of NbAs single crystals was reduced to $\sim$20 $\mu$m by mechanically polishing. Four Pt electrodes were attached to the sample corners with silver epoxy. Resistance measurement under ambient pressure was performed in a Physical Property Measurement System (PPMS, Quantum Design). For high-pressure resistance measurements, the sample was loaded into the chamber together with ruby powder for pressure determining. The silicon oil was used as the pressure-transmitting medium. The high-pressure resistance measurements were performed in a $^3$He cryostat, by the Van der Pauw method. All the samples used in this work were from the same batch.

We first performed the high-pressure powder XRD measurements to examine the structural stability of NbAs. Figure 1(a) displays the XRD patterns of NbAs under various pressures at room temperature. Under low pressure $p$ = 0.47 GPa, all the peaks can be indexed with the noncentrosymmetric tetragonal space group $I4_1md$, and the lattice parameters $a$ = 3.453 {\AA} and $c$ = 11.677 {\AA} are obtained by fitting the XRD data with GSAS software. These values agree well with previous reports under ambient pressure \cite{Bollerh,SaparovB}. With increasing pressure, one can see that the overall XRD pattern does not change. The major (112) peak only shows a slight shift towards higher angle, as plotted in Fig. 1(b). This shift indicates the shrink of lattice under pressure. Under the highest pressure $p$ = 26 GPa we applied in this study, the lattice parameters decrease by 3$\sim$4\%, to $a$ = 3.366 {\AA} and $c$ = 11.206 {\AA}. These results show that the crystal structure of WSM NbAs is very stable under pressure.

Figure 2(a) shows temperature dependence of resistance $R(T)$ for NbAs single crystal (sample A) measured at ambient pressure. The resistance decreases with temperature down to 2 K, with the residual resistance ratio RRR = $R$(300K)/$R$(2K) $\approx$ 6. This metallic behavior is consistent with that of previously reported polycrystalline sample (RRR $\approx$ 2) \cite{SaparovB}. In Fig. 2(b), the low-temperature resistance of sample B under various pressures up to 20 GPa are plotted. With increasing pressure, the $R(T)$ remains metallic, but its absolute value increases monotonically. A very low temperature, there is a small resistance upturn under several pressures ($p$ = 1.5, 16.3, and 19.9 GPa), the origin of which is not clear to us. Nevertheless, no superconducting transition is observed down to 0.3 K under pressure up to 20 GPa. The little change of $R(T)$ behavior and the absence of superconductivity demonstrate that the WSM state in NbAs is very robust under pressure.

Therefore, no dramatic pressure effects are observed on both crystal structure and electronic state of WSM NbAs. This situation is quite different from that of 3D TIs and DSM Cd$_3$As$_2$ \cite{JLZhang,CZhang,Paglione,Matsubayashi,JZhu,PPKong,LLSun,He}. For example, the TI Bi$_2$Se$_3$ shows a structural phase transition from rhombohedral ($R$-3$m$) to monoclinic ($C2/m$) structure near 10 GPa, and pressure-induced superconductivity was observed above 13 GPa \cite{Paglione}. Structural phase transition and superconductivity were also reported in other three TIs ( Bi$_2$Te$_3$, Sb$_2$Se$_3$, and Sb$_2$Te$_3$) under pressure \cite{JLZhang,CZhang,Matsubayashi,JZhu,PPKong}. For the 3D DSM Cd$_3$As$_2$, it undergoes a structural phase transition from tetragonal ($I4_1/acd$) to monoclinic ($P2_1/c$) structure near 3 GPa \cite{LLSun}, and pressure-induced superconductivity was reported above 8 GPa \cite{He}. In this context, the robustness of the WSM state in NbAs may relate to its stable crystal structure.

In summary, we investigate the pressure effects on the crystal structure and electronic state of the WSM NbAs. The powder XRD results show that the crystal structure is stable up to 26 GPa. No resistive superconducting transition was observed down to 0.3 K under pressure up to 20 GPa. It is concluded that the Weyl semimetal state is robust in NbAs, and applying pressure is not a good way to get a topological superconductor from a Weyl semimetal.

We thank Xi Dai and Hongming Weng for their valuable comments and suggestions. This work is supported by the Natural Science Foundation of China, the Ministry of Science and Technology of China (National Basic Research Program No: 2012CB821402 and 2015CB921401), China Postdoctoral Science Foundation No: 2014M560288, Program for Professor of Special Appointment (Eastern Scholar) at Shanghai Institutions of Higher Learning, and STCSM of China (No. 15XD1500200). \\

$^*$ jkdong@fudan.edu.cn

$^\dag$ yangwg@hpstar.ac.cn

$^\ddag$ shiyan$\_$li@fudan.edu.cn

\end{document}